\documentclass[preprint,prl,amsmath,amssymb]{revtex4}

\usepackage{graphicx}
\usepackage{dcolumn}
\usepackage{bm}

\begin{document}

\title{Entanglement crossover close to a quantum critical point}

\author{Luigi Amico and Dario Patan\`e}

\affiliation{ MATIS-INFM $\&$ Dipartimento di Metodologie Fisiche e
    Chimiche (DMFCI), Universit\'a di Catania, viale A. Doria 6,
95125 Catania, ITALY}

\begin{abstract}
We discuss the thermal entanglement close 
to a quantum phase transition by analyzing the pairwise entanglement for one 
dimensional models in the quantum Ising universality class. 
We demonstrate that the entanglement sensitivity to thermal and to quantum fluctuations obeys universal $T\neq 0$--scaling laws. 
We show that the entanglement  exhibits   a peculiar universal crossover behaviour , shedding light on the 
mechanisms bringing quantum criticality up to finite temperatures.
\end{abstract}

\maketitle

Quantum phase transitions (QPTs) are  zero temperature transitions  
driven  by quantum fluctuations\cite{SACHDEV-BOOK}.  Even though absolute zero is not accessible experimentally, 
QPTs have  profound influences at finite temperature, making the phenomenon  detectable\cite{JJA,QHE,HIGH-TC,HEAVY}.
It is  the relative role of the 
thermal over quantum fluctuations inherited from $T=0$ that constitutes  a 
characterization of the system at low temperature\cite{CHN}.
The interplay between  
classical statistical mechanics and quantum mechanics may be  so tight that 
a qualitatively new  phenomenology can  emerge. 
Thereby it is expected that the entanglement could play an important role,  
complementing  traditional approaches employed in statistical 
mechanics\cite{NATURE,OSBORNE,KOREPIN,FIRENZE1,FIRENZE2,DELGADO,VIDAL,CALABRESE,HUBBARD}. 
Though entanglement is generically present in quantum many particle systems, the key to distill it 
quantitatively out of the correlations was provided  only recently  setting the so called ``entanglement measures'', assigning  a precise number to the 
entanglement encoded in a given state\cite{MEASURES,WOOTTERS}.  
Analyzing the entanglement in systems  of many quantum particles now constitutes a new angle of research in the physical literature\cite{ENT-T,VEDRAL}. 
In the extreme case of $T=0$ it  was shown that entanglement and correlation functions share the property 
to be sensitive to the qualitative change of the ground state at the quantum critical point, and 
with the same non-analytic behaviour; however, two-points entanglement and correlation functions are distinct in that the first can be short-ranged whenever the 
system is correlated to all scales\cite{NATURE}.    
Many aspects of QPT were investigated under this light. Various QPT in spin-- $1/2$ models have 
been studied\cite{KOREPIN,FIRENZE1,FIRENZE2}; the string order parameter in gaped spin $1$ systems has been interpreted in terms 
of localizable entanglement\cite{DELGADO}. Connection of the entanglement entropy with  Conformal Field Theory  has been formulated\cite{VIDAL,CALABRESE}. Finally entanglement in 
QPT of itinerant electrons was proved to be related to ground state charge and spin susceptibilities\cite{HUBBARD}. 

Studies on entanglement close to  QPT are confined to zero temperature.
The aim of the present study is to explore quantitatively the role of 
the temperature on the quantum critical properties of entanglement encoded in the state of the system.
We  refer to the finite temperature entanglement and not to ordinary correlation
functions (containing both classical and quantum correlations) to describe the effects fanning out from 
the quantum phase transition, that is a pure quantum phenomenon. We demonstrate
that the  entanglement sensitivity to thermal and to quantum fluctuations
obeys universal $T\neq 0$--scaling laws. 
We show that  the entanglement, together with its
criticality, exhibits  a  crossover behaviour with new aspects
of  it, 
which have universal features within the class of models we considered.
  Finally we elaborate on how typical is the crossover for the 
entanglement.

Concretely,  we focus on a spin chain whose  Hamiltonian is\cite{LIEB}
\begin{equation}
H=-J\sum_{i=1}^N (1+\gamma)S^x_i S^x_{i+1}+
(1-\gamma)S^y_i S^y_{i+1} - h\sum_{i=1}^N S^z_i. \label{model}
\end{equation}
The spins $S^{\alpha}=\frac{1}{2}\sigma^{\alpha}$, $\alpha=x,y,z$	
($\sigma^{\alpha}$ are Pauli matrices) experience an exchange
interaction with coupling $J>0$ and uniform magnetic field of strength
$h$. Besides to model actual compounds\cite{ISING-EXP},
Eq.(\ref{model}) constitutes a standard arena in statistical mechanics
for reliable studies in the theory of quantum critical phenomena.  In
fact at $T=0$, for $N\rightarrow \infty$ the systems (\ref{model})
have a quantum critical point at $a\doteq h/2J=a_c=1$, and for
$0<\gamma \le 1$ they identify the quantum-Ising universality
class. The low temperature behavior is influenced by the interplay
between thermal and quantum fluctuations of the order parameter $\langle S^x\rangle$.
 The renormalized--classical
crosses-over the quantum disordered regimes through the so called
``quantum critical region''\cite{SACHDEV-BOOK,CHN,SACHDEV-ISING}.  In
the $T-a$ plane a $V$-shaped phase diagram emerges, characterized by
the cross-over temperature customarily defined as $T_{cross}\doteq |a-a_c|^{z\nu}$ (the
critical exponents $\nu=z=1$ for the model (\ref{model})) fixing the
energy scale \cite{CONTINENTINO}.  For $T\ll T_{cross}$ the thermal De
Broglie length $\lambda_{th}$ is much smaller than the average spacing
of the excitations $\xi_c$; therefore the correlation functions
factorize in two contributions coming from quantum and thermal
fluctuations separately.  The quantum critical region is characterized
by $T\gg T_{cross}$.  In there $\lambda_{th}\sim \xi_c$, and the
factorization of the correlation functions does not occur.  In this
regime the interplay between quantum and thermal effects is the
dominant phenomenon affecting the physical behaviour of the system.
Here we describe how such interplay is manifested by the finite
temperature two-spins entanglement. We consider the entanglement
between two spins, at chain positions $i$ and $j$.  This can be
expressed by the concurrence\cite{WOOTTERS} $C(R)=\max \{ 0,
\lambda_0(R)-\lambda_1(R)-\lambda_2(R)-\lambda_3(R) \}$, where
$\lambda_n$ are the square roots of the eigenvalues of the matrix
$\sigma_y\otimes \sigma_y \rho^{*} \sigma_y\otimes \sigma_y \rho$, in
descending order, being $\rho$ the reduced density matrix and
$R\doteq |i-j|$. We point out that for the models (\ref{model}),
$C(R)=C(R,a,T)$ can be obtained analytically at any $T$ and $a$
since  $\rho$ is accessible exactly by using results of the
Refs.\cite{McCOY,PFEUTY}. For generic $\gamma $'s, close enough to the
quantum critical point, the range of thermal concurrence results to be
constant; deviations are ultimately due to the factorization of the
ground state at $a=a_f\doteq\sqrt{1-\gamma^2}$ that turns out a non
critical effect\cite{FIRENZE1,FIRENZE2,KURMAN,FACT}. We remark that the
customary analysis\cite{SACHDEV-BOOK,CHN,SACHDEV-ISING} based on the
correlation length and the asymptotics of the correlation
functions is not feasible here, since the 
entanglement extends over a short range,  that is found non-universal\cite{NATURE}. 
Despite of this, we shall evidence how the crossover is reflected at short-distance, on the
entanglement.

The  first aspect  we consider is the scaling of the entanglement close to the quantum critical point.
It was demonstrated that the trait of the QPT can be  captured by the field derivatives $\partial_a C(R)$  diverging
logarithmically at T=0 and $a\rightarrow a_c$, with scaling behaviour\cite{NATURE}. 
At finite temperature we study both  $\partial_a C$ and  $\partial_T C$. 
Following the standard paradigm (see f.i. Refs.\cite{SACHDEV-BOOK,VOJTA})  
the parameters $a$ and $T$ are indicative for the 'strenght' of  thermal and quantum fluctuations 
of the order parameter  respectively.   
Then $\partial_a C$ and $\partial_T C$ can be interpreted  as the response of  the entanglement  
when those fluctuations are  tuned.
Here we analyze the scaling properties of such  two quantities
in a finite temperature neighborhood of the quantum critical point.
We first consider  the Ising model  ($\gamma=1$). Also at $T\neq 0$
$C(1)$ and $C(2)$ are the non vanishing concurrence.

\begin{figure}[ht]\centering
\includegraphics[width=10cm]{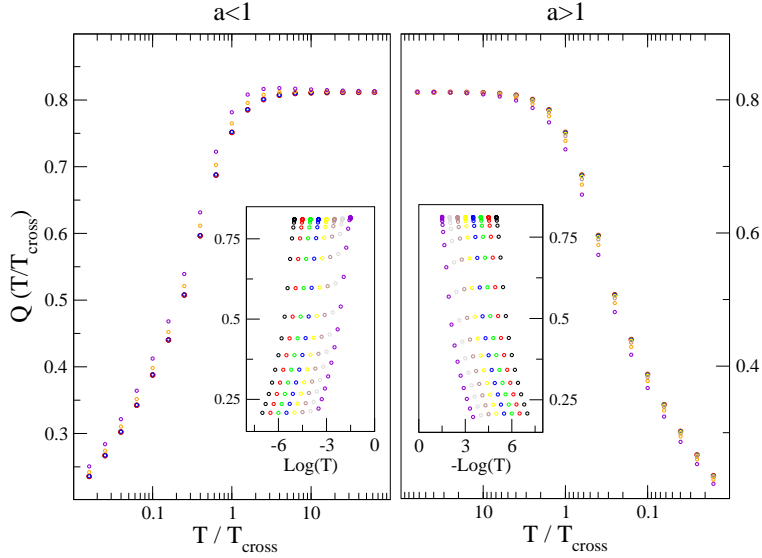}
\caption{Scaling of entanglement at finite temperature close to the quantum critical point. The scaling function 
$Q(T/T_{cross})$ for $\partial_a C(1)$ is presented. Temperatures are expressed in units of $J$.
 $Q=exp(\partial_a C(1))/T^\Upsilon $ as function of the non-scaled coordinate $T$ is  shown in the insets; different colors correspond to different distances $d=\sqrt{(a-a_c)^2+T^2}$  from the quantum critical point.
Plotted data belong to  a neighborhood of the QCP of radius  $10^{-2}\le d \le10^{-5}$(respectively violet-black data);
data closer to the critical point  collapse into the same line asymptotically, thus proving the scaling ansatz 
(the deviations are corrections to the scaling due to  inspections at 
larger distances to the critical point).  
The functional form of $Q$ is obtained by 
imposing that  $\partial_a C(1)$ reduces to the $T=0$ result\cite{NATURE}; for that:  $Q(T\rightarrow 0)\approx \alpha(T/T_{cross})^{-\Upsilon} $;  
Eq.(\ref{ansatz}) reduces to $A\ln|a-a_c|+B$, 
fixing  $B=-\ln\alpha,~A=z\nu  \Upsilon$; $\nu$ e $z$ are related to $\Upsilon$ at  $a=a_c$ since 
$ 
\partial_a|_{a=a_c} C(1)=\Upsilon\ln T+cost
$,  
implying  that $z\nu= A/ \Upsilon$. We found  $\partial_a|_{a=a_c} C(1)=-0.2701 \ln(T)-0.2089$ consistently 
with the critical exponents and the results of \cite{NATURE} . } 
\label{SCALINGNN}
\end{figure}

\begin{figure}[ht]\centering
\includegraphics[width=10cm]{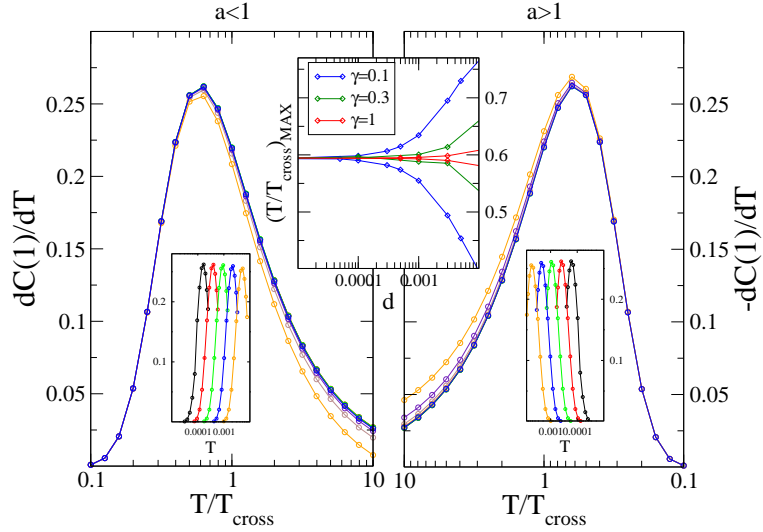}
\caption{The scaling properties of $\partial_T C(1)$ ($\gamma=1$) are verified  for 
$10^{-2}\le d \le 10^{-5}$ around the quantum critical point. 
Such properties are found to be universal and independent of $R$. 
The middle  inset evidences that the maxima are independent of $\gamma$ ; the corresponding  temperatures 
are plotted as functions of the distance to the critical point $d$, for different values of $\gamma$. 
At small $d$ the maximum
 position saturates at $T^*=\alpha T_{cross}$ with 
$\alpha\sim0.595 \pm 0.005$.}
\label{T-derivative}
\end{figure}

For $\partial_a C(R)$, at $T\neq 0$ the divergence at the critical point
is rounded off to  maxima scaling as $a_m\sim a_c -T^{1.72}$, 
that are the finite temperature effects remaining out of  the QPT.
According to the theory of the quantum critical phenomena\cite{QHE,CONTINENTINO,BARBER}, the scaling ansatz is
\begin{equation}\label{ansatz}
\partial_a C(1)\approx\ln[T^{\Upsilon} Q(\frac{T}{T_{cross}})]
\end{equation}
where $\Upsilon$ is a  
non universal exponent; 
Eq.(\ref{ansatz}) was  verified in Fig.\ref{SCALINGNN} where the data collapse of 
$Q$ as function of $T/T_{cross}$ is presented. 
Universality in the critical entanglement was  verified by analyzing (\ref{ansatz}) for different $\gamma$'s (the results are not shown here).  $C(2)$ shows the same universal behaviour.  
\\
Though $\partial_T C(R)$ 
is not diverging at the QPT
it is affected by the quantum criticality. In fact we observe  that
\begin{equation}
\partial_T C(1)\approx  P(\frac{T}{T_{cross}})
\label{ansatz-T}
\end{equation} 
The scaling function  $|P({T}/{T_{cross}})|$ is characterized by pronounced  maxima at certain crossover temperature $T^*=\alpha T_{cross}$ (see the caption of Fig.\ref{T-derivative}). Close enough to the quantum critical point the 
 position of the maxima is found to be 
independent of $\gamma$ and $R$.  
Therefore, by gaiting into the  quantum critical region the   
entanglement crosses-over at $T=T^*$, experiencing the largest variation 
for thermal fluctuations. This effectively ``separates'' the entanglement in the renormalized classical and disordered regimes from that one in the  
quantum critical regions. However, as we shall point out below, such separation is much more involved. 

\begin{figure}[ht]\centering
\includegraphics[width=8cm]{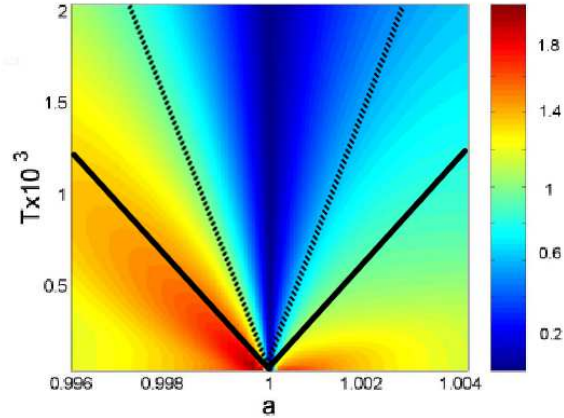}
\caption{Symmetry of the entanglement in the $T-a$ plane.
The directional derivative
$D_{\bf u} C(R)$, 
with $\tan\alpha=T/T_{cross}$.
We observe that the critical point can be  approached with constant  $D_{\bf u} C(R)$ following cusp-like trajectories.  
The density plot refers to the Ising model $\gamma=1$ and to nearest neighbour concurrence $C(1)$.  $T=T^*$ and $T=T_M$  are drawn as dashed and thick lines respectively.  
The same behaviour of the entanglement arises at  
any values of $\gamma$ and $R$.}
\label{dir}
\end{figure}

To study the symmetry of the entanglement in the $a-T$ plane 
we analyzed  the directional derivative 
$D_{\bf u} C(R)\doteq |\nabla C(R) \cdot {\bf u}|=|\partial_T C(R) \sin \alpha+ \partial_a C(R) \cos \alpha|$ as function of  the slopes ${\bf u}\equiv \left (\cos\alpha, \sin \alpha \right )$, $\tan\alpha=T/T_{cross}$  
emerging from the  quantum critical point 
(see the caption of  Fig.\ref{dir}).  
The behavior of $\partial_a C(R)$  and $\partial_T C(R)$ implies  
that  $D_{\bf u} C(R)$ is very small along the critical line 
$a=a_c$, within a cusp-like domain.     
In a sense  within such a region, close to the quantum critical point,  
the pairwise entanglement is particularly ``rigid''. 
 This could provide a mechanism explaining how  
quantum correlations  are robust up to  finite temperatures 
particularly along slopes within the quantum 
critical region. 

\begin{figure}[ht]\centering
\includegraphics[width=8cm]{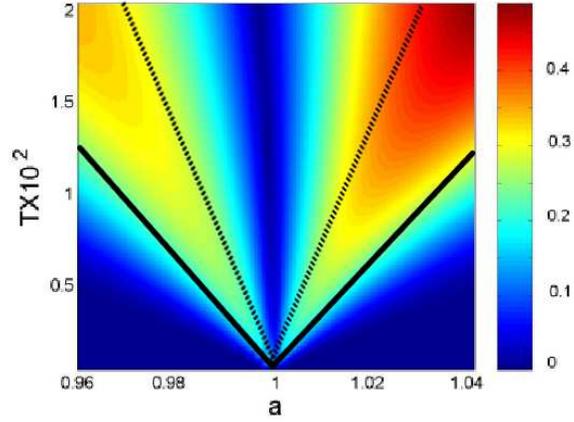}
\caption{ The relative role of thermal over quantum fluctuations of the order parameter on the entanglement 
$\displaystyle{{\partial_T C(1)\over\partial_a C(1)}}$ for $\gamma=1$. The crossover regions 
between the low temperature and quantum critical regions 
are approximately enclosed in  $T_M\lesssim T\lesssim T^*$ which are drawn as thick and dashed lines respectively.
We notice that such a ratio has also a geometrical meaning in that 
it can be obtained as the  angle  of steepest concurrence variation:
$\partial_\alpha [D_{\bf u} C(R) ]=0$. Therefore the two regions where $\displaystyle{\partial_a C(1)}$  is  dominant  correspond to  abrupt concurrence 
changes  at $\alpha_s=0$.
line).}
\label{ratio}
\end{figure}

\begin{figure}[ht]\centering
\includegraphics[width=8cm]{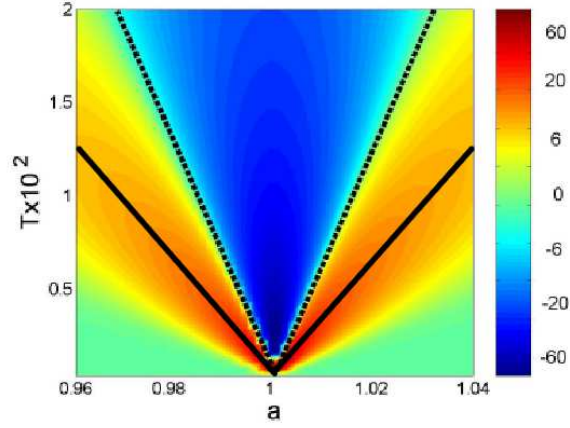}
\caption{The effect of temperature on the anomalies originated from 
the critical divergence of the field-derivative of $C(R)$ can be measured by 
$\partial_T[\partial_{a} C(R)]$. The density plot  corresponds to $\gamma=1$ and 
$R=1$. $T=T^*$ and $T=T_M$  are drawn as dashed and thick lines respectively. Maxima below $T^*$ are found  at $T_M =\beta  T_{cross}$ 
with $\beta\sim 0.290 \pm 0.005$ and they are independent of $\gamma$ and $R$; the crossover behaviour is enclosed in 
between  the two flexes of  $\partial_T[\partial_{a} C(R)]$ at  
 $T_{c1}$ $T_{c2}$; such values are fixed to: $T_{c1}=(0.170  \pm 0.005) T_{cross}$ and 
$T_{c2}=(0.442  \pm 0.005)T_{cross}$ and 
found to be independent of $\gamma$ and $R$. For $T\lesssim T_{c1}$ $\partial_T[\partial_{a} C(R)]\simeq 0$.
Scaling properties are inherited in $\partial_T[\partial_{a} C(R)]$  from $\partial_{a} C(R)]$ (see Eq.(\ref{ansatz})).}
\label{mixed}
\end{figure}

Now we analyze  how the entanglement is affected by  the 
interplay between quantum and thermal effects. 
 We consider the quantity: $\partial_T C(R)/\partial_a C(R)$  
(we note that a formally similar quantitity, the so called Grueneisen 
parameter, emerged in recent studies of  quantum criticality\cite{GRUENSEN}).
This quantity can be 
 interpreted as  a  measure of the relative role  of the effects of thermal and 
quantum   fluctuations of the order parameter on the entanglement. 
The analysis evidences two 
regimes (Fig.\ref{ratio}).
$\partial_a C(R)$  is dominant at very low 
temperature and near  the critical line $a=a_c$. Instead in the  crossover  to 
the quantum disordered and renormalized classical regimes the role of 
entanglement's thermal sensitivity is enhanced.
The importance of the effect of temperature in the entanglement crossover 
can be pointed out also analyzing  how the critical divergence of $\partial_{a} C(R)$ 
is affected by  variations of the temperature: 
$\partial_T[\partial_{a} C(R)]$. 
Due to the vanishing of the gap at the quantum critical point,  
in the region $T\gg T_{cross}$
an arbitrarily small temperature 
is immediately effective  in the system.
Interestingly enough  we observe that  
$\partial_T[\partial_{a} C(R)]$is enhanced also 
along the line $T_M = \beta 
T_{cross}$ with $\beta<\alpha$  and 
spread out on a finite range of $T/T_{cross}$: $T_{c1}\lesssim T \lesssim T_{c2}$ (see the caption of Fig.\ref{mixed}); the 
values of  $T_M$, $T_{c1}$ and $T_{c2}$ are proved to be independent of $\gamma$ and $R$. We notice 
that  $T_M$  characterizes also the crossover in Fig.\ref{ratio} that is approximately enclosed between $T^*$ and $T_M$. 
The analysis of the signs of $\partial_T[\partial_{a} C(R)]$ leads 
to the conclusion that for $T > T^*$ a reduction of the temperature 
 corresponds to an increase of the effects of the critical 
behavior of $\partial_{a} C(R)$; while for  $T < T^*$ a reduction of the temperature  effectively leads  the system away from  the criticality, 
consistently with the  important role of the energy gap scale in such a regime\cite{SACHDEV-BOOK,CONTINENTINO}.  For  $T\lesssim T_{c1}$ the temperature 
does not affect $\partial_a C(1)$.

To estimate the impact of the entanglement  in the total correlation   we 
apply a variation of the arguments developed in \cite{ANFOSSI}. Specifically, we 
analyze  the quantum mutual information amoung the $i-th$ and $j-th$ spin: 
$I_{ij}=S_i+S_j-S_{ij}$, with $S_x=Tr\rho_x\log \rho_x$ being the Von Neumann entropy of 
$\rho_x$\cite{VEDRAL-MUTUAL}, as a valuable measure of {\it quantum $\&$ classical} correlations, 
taking into account of all two-point spin correlation functions.
It emerges that a similar  phenomenology found for the concurrence is  experienced by $I_{ij}$; 
moreover we found that the crossover temperatures for $I_{ij}$  are  shifted by less than $10\%$ respect 
to the corresponding temperatures for the the entanglement. Then: By distilling out the quantum part from the total correlation, it 
emerges how the crossover is little affected by the classical correlations,  
suggesting that it is driven solely by the interplay between  thermal and quantum  sensitivities of the  entanglement .   

The present analysis reveals that the  entanglement sensitivities 
to thermal and quantum fluctuations  close to a QPT 
are universal functions of the angle at which the critical point is 
approached.  
By studying how the entanglement in
the ground state of the system is affected by the contribution coming
from the excitations, we analyzed how \textit{pure quantum effects} are
affected by the finite temperature.  In particular, it is shown that the entanglement
exhibits a crossover behaviour evidencing a  ``fine structure'',  characterized by   energies $T^*$ and $T_M$ that are found independent of
 the specific model in the class (\ref{model}).  
Interestingly enough, a similar anomaly in the entanglement sensitivity was 
evidenced recently in the context of the spin-boson model. In fact it  was proved 
that the coherent-incoherent oscillations crossover is  accompained  by the largest variation of the entanglement between the two-level system and the bosonic bath\cite{Stauber} (see also~\cite{Kopp-ent}). 

Our scaling result together with the study of the entanglement crossover  are particularly
significant, in our opinion, in that they involve pure quantum correlations at a short-range; this is pointed out by analyzing also 
the quantum mutual information.
>From our analysis a neat meaning is sheded on the specific crossover temperature we found,
beyond its standard interpretation  as  a mere reference energy scale for the crossover. 

In statistical mechanics, our analysis can constitute a crucial ingredient  for a deeper understanding of the role of the
quantum distilled part of the correlations entanglement in the crossover
phenomena  close to QPTs.
As drawn in \cite{COLEMAN,KOPP}, spin-off in the research on
the mechanisms behind
the puzzling low temperature behavior of  exotic materials like
high $T_C$ superconductors and heavy-fermions are possible.
 We point out that, being at finite temperature, our study can open the way 
towards an experimental observation of the physical effects that  
entanglement has on the criticality of the system. 
We believe that  the present analysis could be 
important  both for pure research and quantum computation perspectives.

{\bf Acknowledgments}.
We thank  G. Falci,  R. Fazio, A. Fubini, A. Osterloh, and J. Siewert 
for  support and discussions.

\newpage

\end{document}